\documentclass[12pt,twoside,a4paper]{article} 
\usepackage{amsmath,amssymb,latexsym,theorem,bbm,shapepar,natbib,epsfig}
\newfont{\msa}{msam10 scaled\magstep1}
\newfont{\ssmsa}{msam9}

\def\crps{\mathop{\hbox{\rm crps}}}

\numberwithin{equation}{section}

\title{Probabilistic temperature forecasting with statistical
  calibration in Hungary} 

\author{{\sc S\'andor Baran}\\ 
         Faculty of Informatics, University of Debrecen\\
         Kassai \'ut 26, H-4028 Debrecen, Hungary \\ 
        and \\
        Institute of Applied Mathematics, University of Heidelberg\\
        Im Neuenheimer Feld 294, D-69120 Heidelberg, Germany \\ [3mm]
       {\sc Andr\'as Hor\'anyi} \\   
       Hungarian Meteorological Service \\
       P.O. Box 38, H-1525 Budapest, Hungary \\ [3mm]
       {\sc D\'ora Nemoda}\\ 
         Faculty of Informatics, University of Debrecen\\
         Kassai \'ut 26, H-4028 Debrecen, Hungary 
     }

\date{}

\begin{document}
\pagestyle{myheadings}

\maketitle

\begin{abstract}
Weather forecasting is mostly based on the outputs of
deterministic numerical weather forecasting models.  Multiple runs of
these models with different initial conditions result in
forecast ensembles which is are used for estimating the distribution of future 
atmospheric variables. However, these ensembles are usually
under-dispersive and uncalibrated, so post-processing is required.

In the present work Bayesian Model Averaging (BMA) is applied for
calibrating ensembles of temperature forecasts produced by the
operational Limited Area Model Ensemble Prediction System of the
Hungarian Meteorological Service (HMS). 

We describe two possible BMA models for temperature data of the HMS and
show that BMA post-processing significantly improves calibration and
probabilistic forecasts although the accuracy of point forecasts is
rather unchanged. 

\end{abstract}

\section{Introduction}
   \label{sec:sec1}
The main objective of weather forecasting is to give a robust and
reliable prediction of future atmospheric states on the basis of
observational data, prior forecasts valid for the initial time
of the forecasts and mathematical models describing the dynamical and
physical behaviour of the atmosphere. 
These models numerically solve the set of the hydro-thermodynamic
non-linear partial differential equations of the atmosphere and its
coupled systems (like surface or oceans for instance). The difficulty
with these numerical weather prediction models is that since the
atmosphere has a chaotic character the solutions  
strongly depend on the initial conditions and also on other uncertainties 
related to the numerical weather prediction process. In practice, 
the results of such models are never fully accurate and 
forecast uncertainties should be also taken into account in the forecast
preparation.       
A possible solution is to run the model with different
initial conditions (since the uncertainties in the initial conditions
are one of the most important sources of uncertainty) and produce an
ensemble of forecasts. The forecast ensemble can estimate the
probability distribution of future weather variables which allows
probabilistic weather forecasting \citep{gr}, where not only the
future atmospheric states are 
predicted, but also the related uncertainty information (which is
indeed a valuable extension to the so called deterministic approach,
where only the forecast is given without uncertainty estimation). 
The ensemble
prediction method was proposed by \citet{leith} and since its first
operational implementation \citep{btmp,tk} it has become a widely used
technique all over the world and the users understand more and more
its merits and economic value as well. However, although e.g. the ensemble mean 
on average yields better forecasts of a meteorological quantity than
any of the individual ensemble members, it is often the case that the
ensemble is under-dispersive 
and in this way, uncalibrated \citep{bhtp}, so that calibration is
needed to account for this deficiency.  

The Bayesian Model Averaging (BMA) method for
post-processing ensembles in order to calibrate them was introduced by
\citet{rgbp}. The basic idea of BMA is that for each member of the ensemble
forecast there is a corresponding  conditional probability
density function (PDF) that can be interpreted as the
conditional PDF of the future weather quantity provided the considered
forecast is the best one. Then the BMA predictive PDF of the future
weather quantity is the weighted sum of the individual PDFs
corresponding to the ensemble members and the weights are based on the
relative performance of the ensemble members during a given training
period. In practice, the performance of the individual
ensemble members should have a clear characteristic (and not a random
one) or if it is not the case this fact should be taken into account
at the calibration process \citep[see e.g.][]{frg}.  In \citet{rgbp} the BMA
method was successfully applied to obtain 48 hour forecasts of surface
temperature and sea level pressure in the North American Pacific
Northwest based on the 5 members of the University of Washington
Mesoscale Ensemble \citep{gm}. These weather quantities can be
modeled by normal distributions, so the predictive PDF is a Gaussian mixture.
Later, \citet{srgf} developed a discrete-continuous BMA model for
precipitation 
forecasting, where the discrete part corresponds to the event of no
precipitation, while the cubic root of the precipitation amount
(if it is positive) is modeled by a gamma distribution. In
\citet{sgr10} the BMA method was used for wind speed forecasting and 
the component PDFs follow gamma distributions. Finally, using von Mises
distribution to model angular data 
\citet{bgrgg} introduced a BMA scheme to predict surface wind
direction.

In the present work the BMA method is applied for calibrating ensemble
temperature forecasts produced by the operational Limited
Area Model Ensemble  
Prediction System (LAMEPS) of the Hungarian Meteorological Service
(HMS) called ALADIN-HUNEPS \citep{hagel, horanyi}. ALADIN-HUNEPS
covers a large part of Continental Europe with a horizontal resolution
of 12 km and it is obtained by dynamical downscaling (by the ALADIN
limited area model) of the global
ARPEGE based PEARP system of M\'et\'eo France \citep{hkkr,dljn}. The
ensemble consists of 11 members, 10 initialized from perturbed initial
conditions and one control member from the unperturbed analysis. This
construction implies that the ensemble contains groups of exchangeable
forecasts (the ensemble members cannot be distinguished, thus it is
not possible to depict a systematic behaviour of each member), so for 
post-processing one has to use the modification of 
BMA as suggested by \citet{frg}. We remark, that BMA method has
already been successfully applied for wind speed 
ensemble forecasts of the ALADIN-HUNEPS -- in \citet{bhn} it is
shown that BMA post-processing of these forecasts significantly
improves the calibration and the accuracy
of point forecasts as well. Now this latter work is extended towards
the calibration of the 2m temperature ensemble forecasts.

\section{Data}
  \label{sec:sec2}

As it was mentioned in the introduction, BMA post-processing
of ensemble predictions was applied for temperature data produced by
the operational ALADIN-HUNEPS system  of the HMS. The 
data file contains 11 member ensemble (10 forecasts started from perturbed
initial conditions and one control) of 42 hour 
forecasts for 2m temperature (given in Kelvin)
for 10 major cities in 
Hungary (Miskolc, Szombathely, Gy\H or, Budapest, Debrecen, Ny\'\i regyh\'aza,
Nagykanizsa, P\'ecs, Kecskem\'et, Szeged) together with the corresponding
validating observations, for the period between October 1, 2010 and March 25,
2011. The forecasts are
initialized at 18 UTC. The data set is fairly  
complete, since there are only two days 
(18.10.2010 and 15.02.2011) when three ensemble members are missing
for all sites and one day (20.11.2010) when no forecasts are available. 

\begin{figure}[t]
\begin{center}
\leavevmode
\epsfig{file=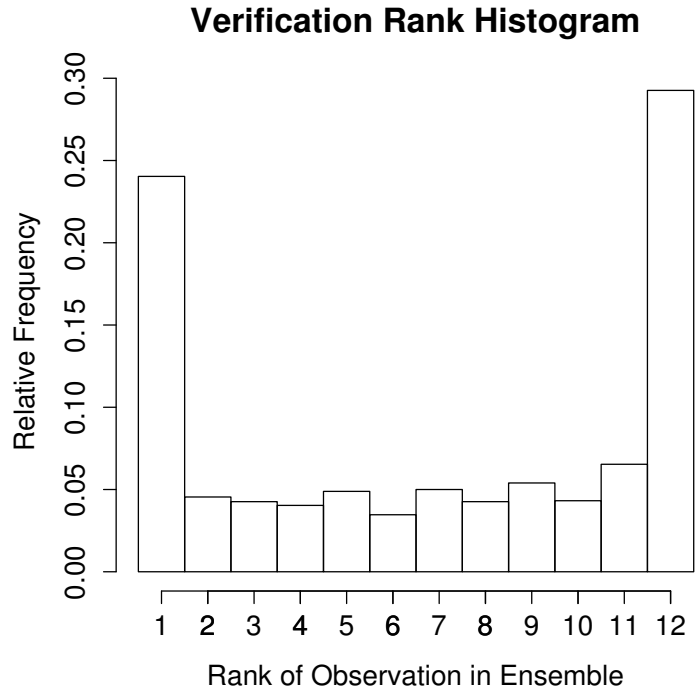,height=9cm}
\caption{Verification rank histogram  for 2m temperature of the
  11-member ALADIN-HUNEPS  
  ensemble. Period: October 1, 2010 -- March 25, 2011.} 
\label{fig:fig1}
\end{center}
\end{figure}

Figure \ref{fig:fig1} shows the verification rank histogram of the raw
ensemble, that is the histogram of ranks of validating
observations with respect to the corresponding ensemble
forecasts \citep[see e.g.][Section 8.7.2]{wilks2}. It is clear that
this histogram is  
far from the desired uniform distribution, in many cases the
ensemble members either underestimate, or overestimate the validating
observations (the ensemble range contains the observed near-surface temperature
values only in $46.36\%$ of the cases).  Hence, the ensemble is under-dispersive
and in this way it is uncalibrated. In case of proper calibration the
probability of a validating observation being below the ensemble range
would be $1/12$ and we would have the same probability for the
observation being above
it. In this way the probability $10/12$ (i.e. $83.33 \,\%$) can be
considered as the nominal value of the ensemble range.

\section{The model and diagnostics}
   \label{sec:sec3}

In order to calibrate the ALADIN-HUNEPS ensemble forecasts for
temperature the modification of   
BMA normal model of \citet{rgbp} for ensembles with
exchangeable members \citep{frg} was used. The grouping of ensemble
members was similar to the case of wind speed investigated in
\citet{bhn}. In the first model we have two exchangeable groups: one contains 
the control denoted by \ $f_c$, \ the other one the 10 ensemble members
corresponding to the different perturbed initial conditions which are denoted by
\ $f_{\ell,1},\ldots ,f_{\ell,10}$, \ respectively. \ We assume
that the probability density function (PDF) of the forecasted
temperature \ $x$ \ equals:  
\begin{align}
  \label{eq:eq3.1}
p\big(x | f_c,f_{\ell,1},\ldots ,
f_{\ell,10};b_{c,0},b_{c,1},b_{\ell,0},b_{\ell,1},\sigma^2; \omega\big)=&\, \omega
g\big(x |f_c,b_{c,0},b_{c,1},\sigma^2\big) \\ &+  \frac 
{1-\omega}{10} \sum_{j=1}^{10}  g\big(x|f_{\ell,j},b_{\ell,
  0},b_{\ell, 1},\sigma^2\big), \nonumber 
\end{align} 
where \ $\omega\in [0,1]$, \ and \ $g(x;f,b_0,b_1,\sigma^2)$ \ is a
normal PDF with mean 
\ $b_0+b_1 f$ \ (linear bias correction) and variance \ $\sigma^2$. \
In this way the 
two groups have different mean parameters \ $b_{c,0}, \ b_{c,1}$ \ and
\ $b_{\ell,0}, \ b_{\ell,1}$, \ and a common standard deviation parameter \
$\sigma$. \ Mean parameters \ \ $b_{c,0}, \ b_{c,1}$ \ and
\ $b_{\ell,0}, \ b_{\ell,1}$ \ are estimated with
linear regression, while the weight parameter \ $\omega$ \
and the variance 
\ $\sigma^2$, \ by maximum likelihood method, using training
data consisting of ensemble members and verifying observations from
the preceding \ $n$ \ days (training period). The maximum of the likelihood
function is found with the  EM algorithm \citep{mclk} and due
to the normality of the model both the expectation (E) and the
maximization (M) step lead to closed formulae which allows fast
calculations. For more 
details see \citet{frg}. Once the estimated 
parameters for a given day are available,  one can
use either the mean or the median of the predictive PDF
\eqref{eq:eq3.1} as a point forecast. 

We also investigate special cases when in model \eqref{eq:eq3.1}
only additive bias correction is present, that is \
$b_{c,1}=b_{\ell,1}=1$, \  and when we do not use bias
correction at all, i.e. \ $b_{c,0}=b_{\ell,0}=0$ \ and \
$b_{c,1}=b_{\ell,1}=1$. 

\begin{figure}[t]
\begin{center}
\leavevmode
\epsfig{file=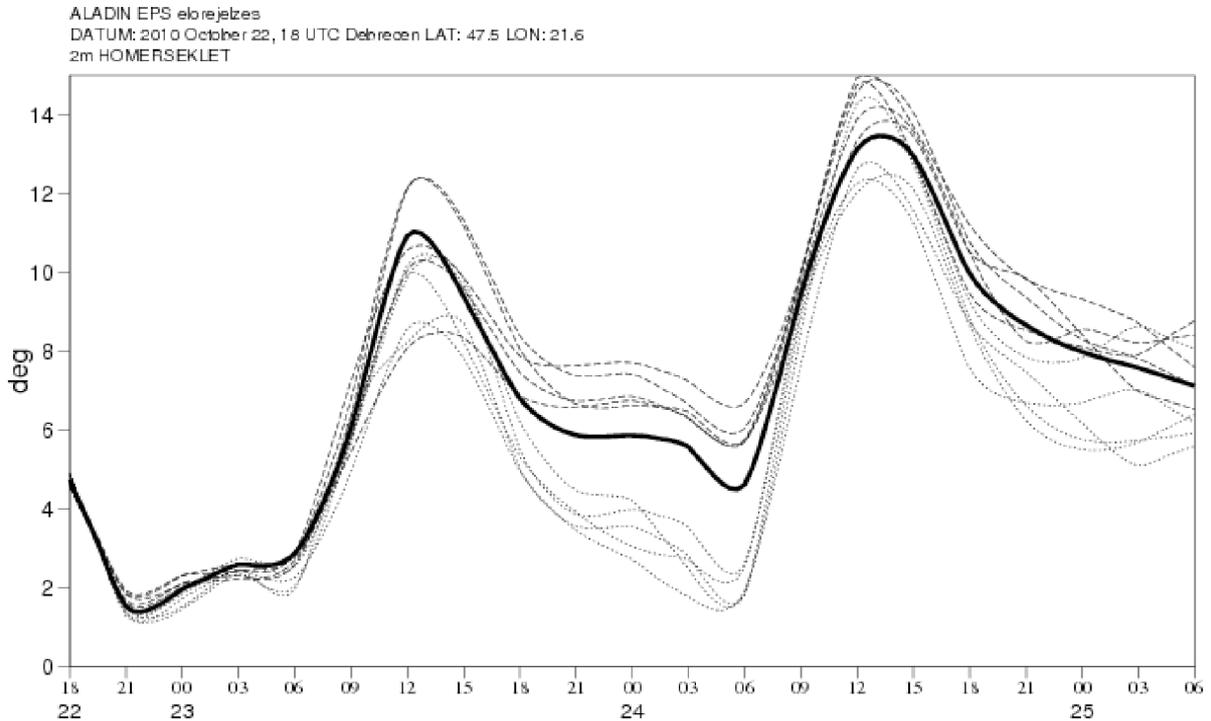,height=10cm}
\caption{Plume diagram of ensemble forecast of 2m temperature
  for Debrecen initialized at 18 UTC, 22.10.2010 (solid line: control;
dotted line: odd numbered members; dashed line: even numbered
members).} 
\label{fig:fig2}
\end{center}
\end{figure}
Now, similarly to wind speed data investigated in \citet{bhn},
to obtain the ten exchangeable ensemble members only five perturbations
are calculated and then they are added to (odd numbered members) and
subtracted from (even numbered members) the unperturbed initial
conditions \citep{horanyi}. Figure \ref{fig:fig2} shows the plume
diagram of ensemble forecasts of 2m temperature
for Debrecen initialized at 18 UTC, 22.10.2010. This diagram clearly
illustrates that the behaviour of 
ensemble member groups \ $\{f_{\ell,1}, \ 
f_{\ell,3}, \ f_{\ell,5}, \ f_{\ell,7}, \ f_{\ell,9}\}$ \ and \ $\{f_{\ell,2}, \
f_{\ell,4}, \ f_{\ell,6}, \ f_{\ell,8}, \ f_{\ell,10}\}$ \  differ
from each other (particularly look at the 24--36h and 48--60h forecast
ranges). Therefore, in this way one can also consider a model with
three exchangeable groups: control, odd numbered exchangeable members and even
numbered exchangeable members. This idea leads to the following PDF
of the forecasted wind speed \ $x$:
\begin{align}
  \label{eq:eq3.2}
q\big(x | f_c,f_{\ell,1},\ldots ,
f_{\ell,10};b_{c,0},b_{c,1},&\,b_{o,0},b_{o,1},b_{e,0},b_{e,1},\sigma^2;
\omega_c,\omega_o,\omega_e\big)= \omega_c
g\big(x|f_c,b_{c,0},b_{c,1},\sigma^2\big) \\ &+ 
\sum_{j=1}^{5} \Big(\omega_o g\big(x|f_{\ell,2j-1},b_{o,0},b_{o,1},\sigma^2\big)+ 
\omega_e g\big(x|f_{\ell,2j},b_{e,0},b_{e,1},\sigma^2\big)\Big), \nonumber 
\end{align} 
where for weights \ $\omega_c,\omega_o,\omega_e\in[0,1]$ \ we have  \
$\omega_c+5\omega_o+5\omega_e=1$, \ while the definition of the PDF \
$g$ \ and of the parameters \ $b_{c,0},\ b_{c,1}, \ b_{o,0}, \
b_{o,1},\ b_{e,0},\ b_{e,1}$ \ and \ $\sigma^2$ \  remains the same as
for the model \eqref{eq:eq3.1}.  Obviously, both the weights and the parameters 
can be estimated in the same way as before.  

\begin{figure}[t!]
\begin{center}
\leavevmode
\hbox{
\epsfig{file=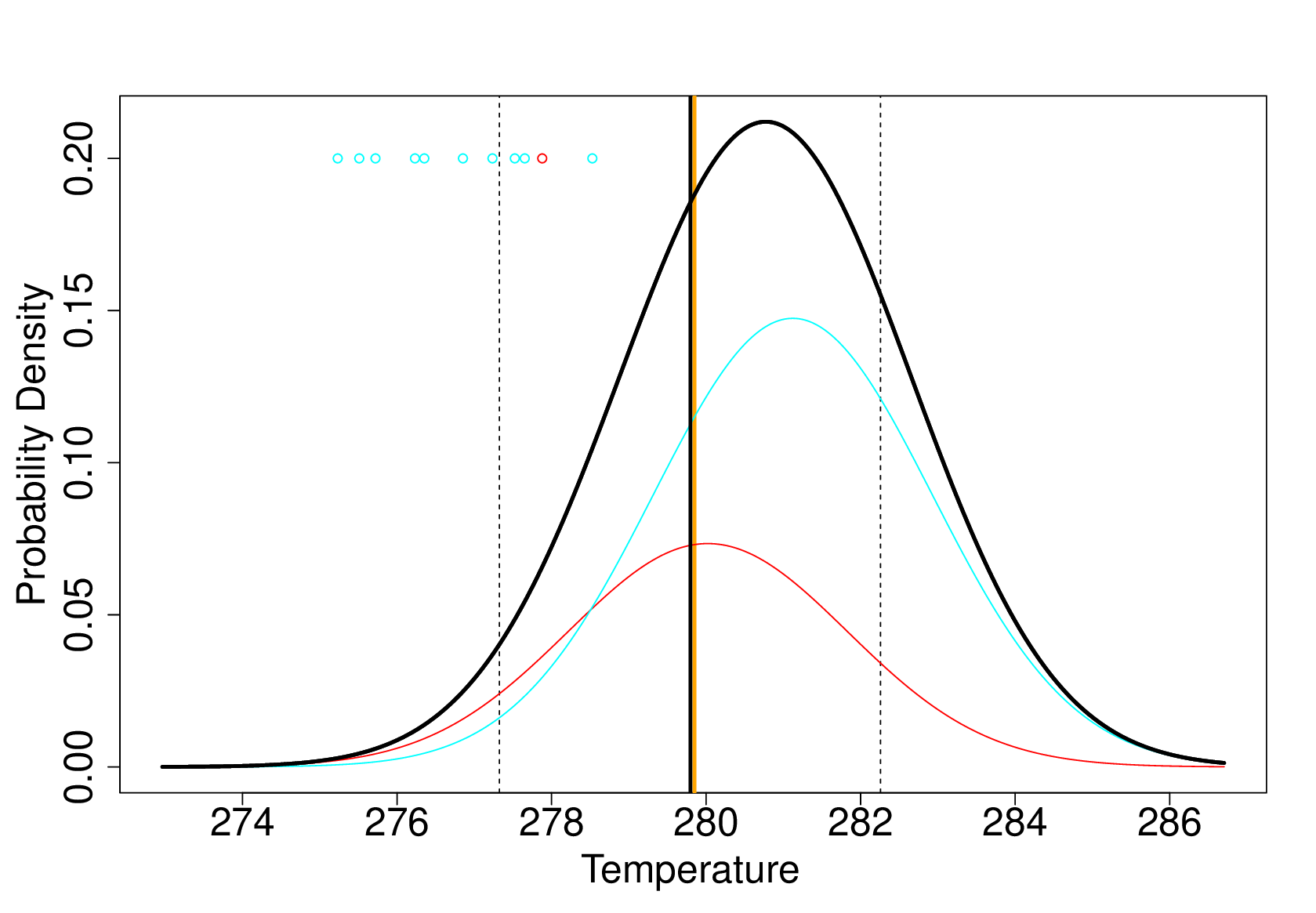,height=8cm, angle=-90} \quad
\epsfig{file=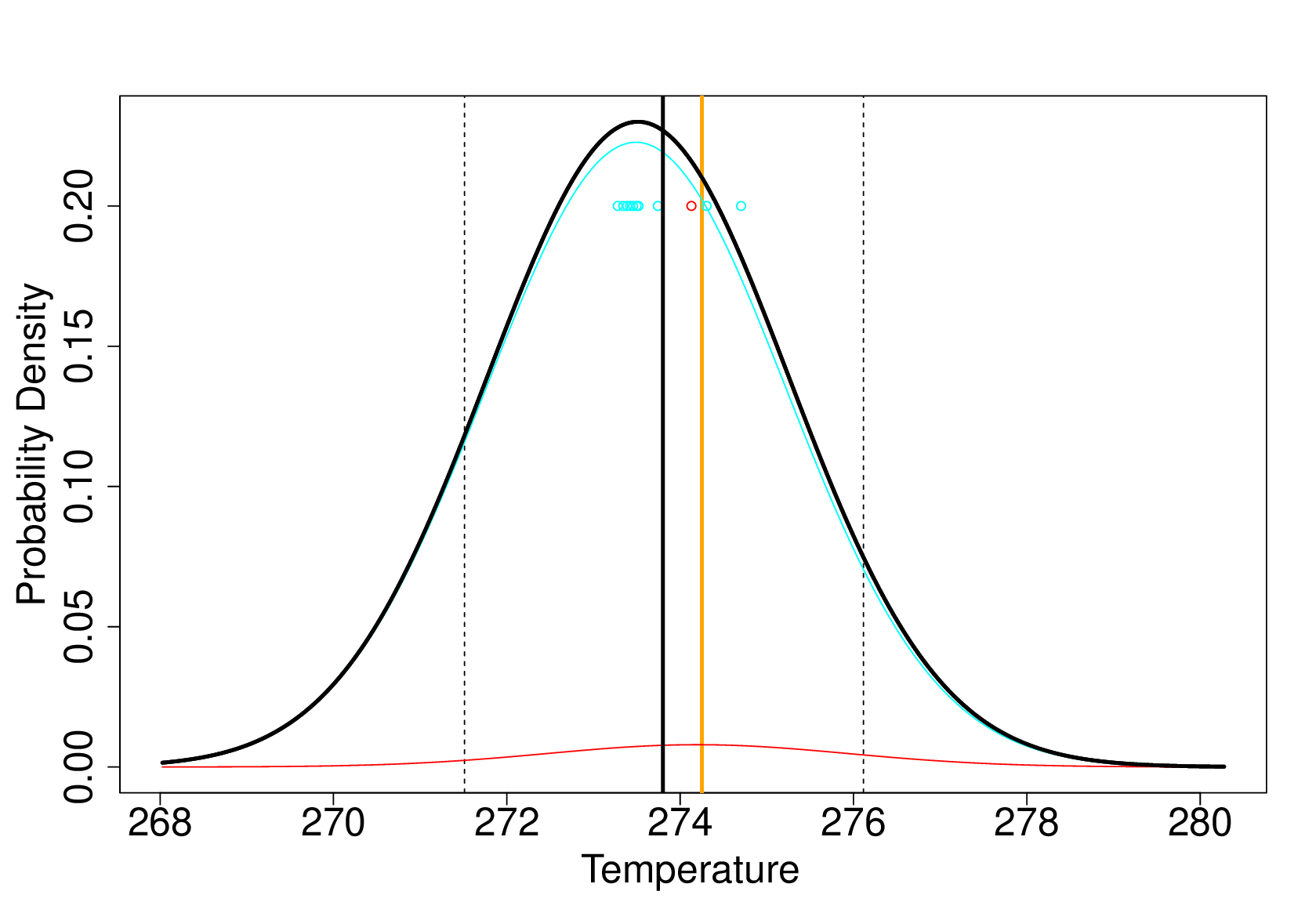,height=8cm, angle=-90}}

\centerline{\hbox to 9 truecm {\scriptsize (a) \hfill (b)}}

\hbox{
\epsfig{file=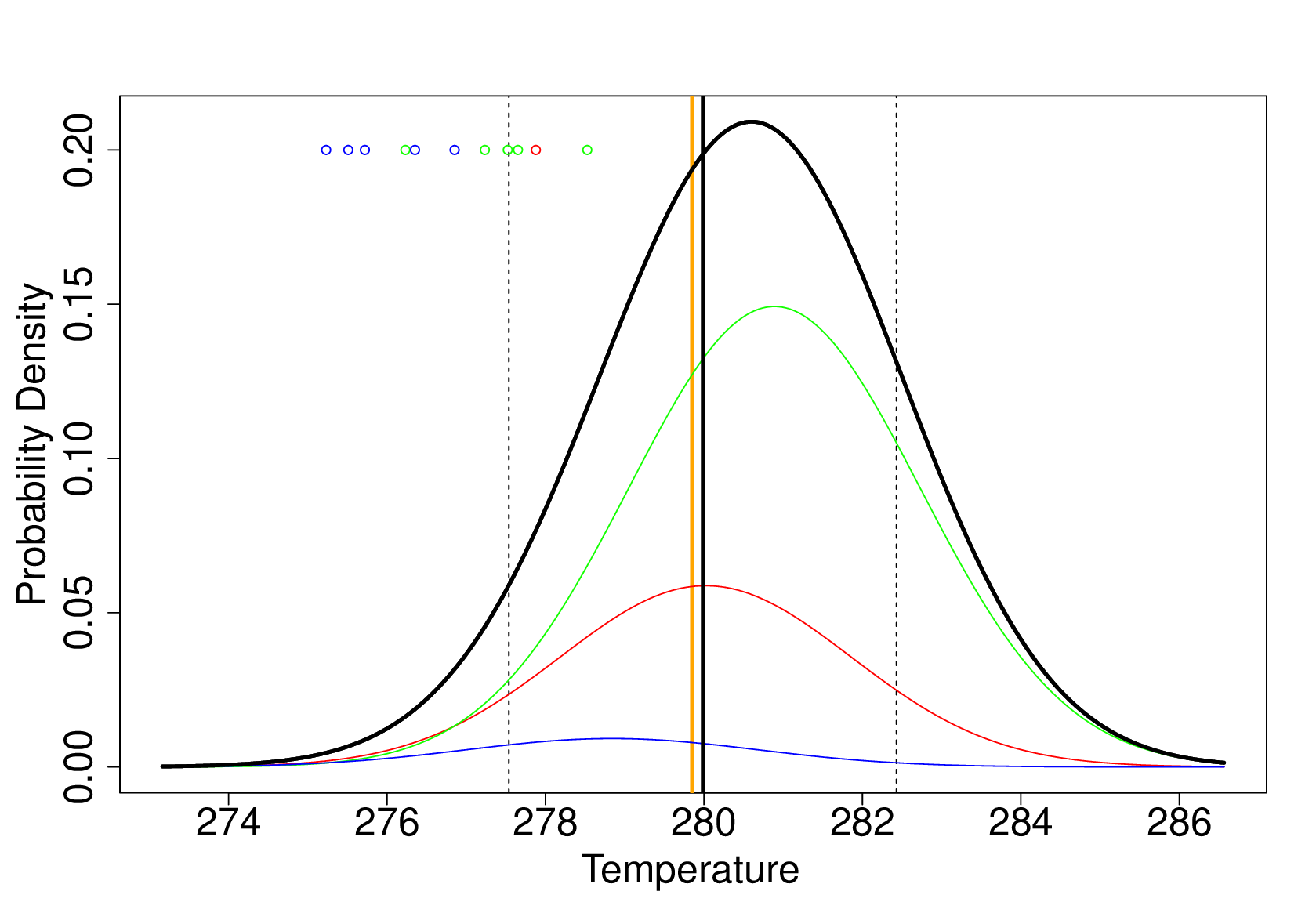,height=8cm, angle=-90} \quad
\epsfig{file=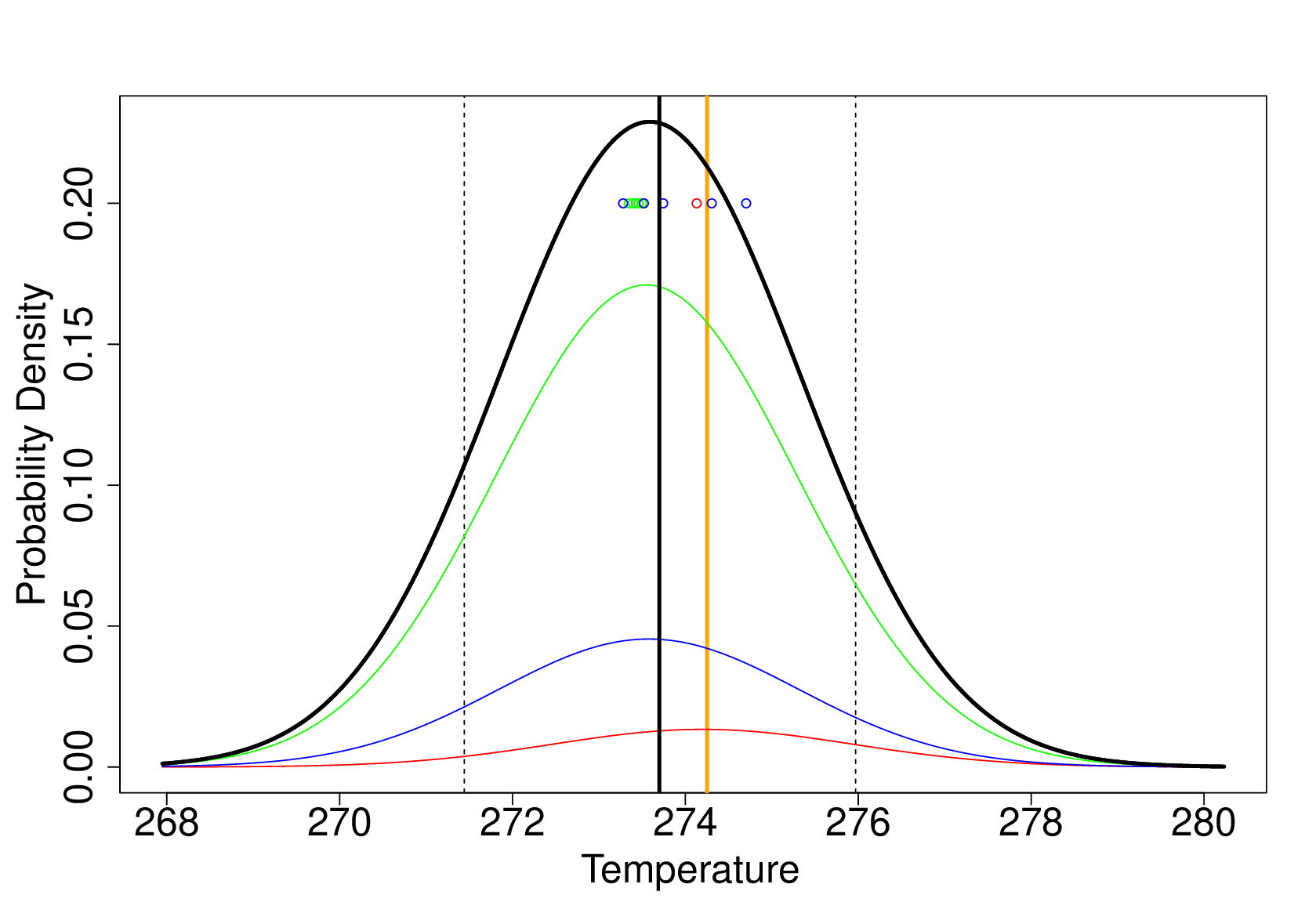,height=8cm, angle=-90}}

\centerline{\hbox to 9 truecm {\scriptsize (c) \hfill (d)}}
\caption{Ensemble BMA PDFs (overall: thick black line; control: red
  line; sum of exchangeable members on (a) and (b): light blue line;
  on (c) and (d): green (odd members) and blue (even members) lines),
  ensemble members (circles with the same colours as the corresponding
  PDFs), ensemble BMA median forecasts (vertical black line),
  verifying observations (vertical orange line) and the first and last
deciles (vertical dashed lines) for 2m temperature in Debrecen for
  models  \eqref{eq:eq3.1}: (a) 24.11.2010, (b) 21.12.2010; and
  \eqref{eq:eq3.2}: (c) 24.11.2010, (d) 21.12.2010; with linear bias
  correction.}  
\label{fig:fig3}
\end{center}
\end{figure}

Similarly to the two-group case, we also investigate model
\eqref{eq:eq3.2} with additive bias correction \
($b_{c,1}=b_{o,1}=b_{e,1}=1$) \ and without bias correction at all \
($b_{c,0}=b_{o,0}=b_{e,0}=0$ \ and \ $b_{c,1}=b_{o,1}=b_{e,1}=1$). \
We also remark, that according to our experiments with the current data
set, the use of different variance parameters for the different groups in models
\eqref{eq:eq3.1} and \eqref{eq:eq3.2} does
not lead to a significant improvement in the performance of the
corresponding forecasts. 

As an illustration we consider the data and forecasts for 
Debrecen for two
different dates 24.11.2010 and 21.12.2010 illustrating two typical
situations for models \eqref{eq:eq3.1}
and \eqref{eq:eq3.2} with linear bias correction. 
Figures \ref{fig:fig3}a and \ref{fig:fig3}b show the  PDFs of
the two groups in model \eqref{eq:eq3.1}, the overall 
PDFs, the median forecasts,
the verifying observations, the first and last
deciles  and the ensemble members. The same functions and
quantities can be seen in Figures \ref{fig:fig3}c and \ref{fig:fig3}d,
where in addition to the overall PDF we have the three component PDFs and three
groups of ensemble members. On 24.11.2010 the spread of the ensemble
members is reasonable (the ensemble range equals 3.3 K) but all
ensemble members underestimate the validating
observation (279.9 K). Obviously the same holds for the ensemble
median (276.9 K), while BMA median forecasts corresponding to the two- and
three-group models (279.8 K and 280.0 K, respectively) are
quite close to the true temperature.  A different situation is
illustrated on Figures \ref{fig:fig3}b and \ref{fig:fig3}d, where the
ensemble range is only 1.4 K, but it contains the validating
observation (274.3 K). The ensemble median (273.5 K) slightly underestimates
the true temperature, but the BMA post-processing yields more accurate
estimators also in this case. The median forecasts corresponding to
models \eqref{eq:eq3.1} and \eqref{eq:eq3.2} are 273.8 K and 273.7
K, respectively.  

In order to check the performance of probabilistic forecasts based on
models \eqref{eq:eq3.1} and \eqref{eq:eq3.2} and the
corresponding point forecasts, as a reference we use the ensemble mean
and the ensemble median. We compare the mean absolute error (MAE) and
the root mean square error (RMSE) of
these point forecasts and also the mean continuous ranked probability score
(CRPS) \citep{grjasa,wilks2} and the coverage and average width of
$83.33\,\%$  central prediction intervals of the BMA
predictive probability  
distributions and of the raw ensemble. The coverage of this central
prediction interval allows a direct comparison to the raw ensemble,
where the ensemble of forecasts corresponding to a given location and
time is considered as a 
statistical sample and the sample quantiles are calculated according
to \citet[Definition 7]{hf}.
 We remark that for MAE and RMSE
the optimal point forecasts are the median and the mean, respectively
\citep{gneiting11, pinhag}.  Further, given a cumulative distribution
function (CDF) \ $F(y)$ \ and a real number \ $x$, \ the CRPS is defined as
\begin{equation*}
\crps\big(F,x\big):=\int_{-\infty}^{\infty}\big (F(y)-{\mathbbm 
  1}_{\{y \geq x\}}\big )^2{\mathrm d}y.
\end{equation*}
The mean CRPS of a probability forecast is the average of the CRPS values
of the predictive CDFs and corresponding validating observations taken
over all locations and time points considered. For the raw ensemble
the empirical CDF of the ensemble replaces the predictive CDF. The
coverage of a \ $(1-\alpha)100 \,\%, \ \alpha \in (0,1),$ \ central prediction
interval is the proportion 
of validating observations located between the lower and upper \
$\alpha/2$ \ quantiles of the predictive distribution. For a
calibrated predictive PDF this value should be around \ $(1-\alpha)100
\,\%$.   

\section{Results}
   \label{sec:sec4}

\begin{figure}[t!]
\begin{center}
\leavevmode
\epsfig{file=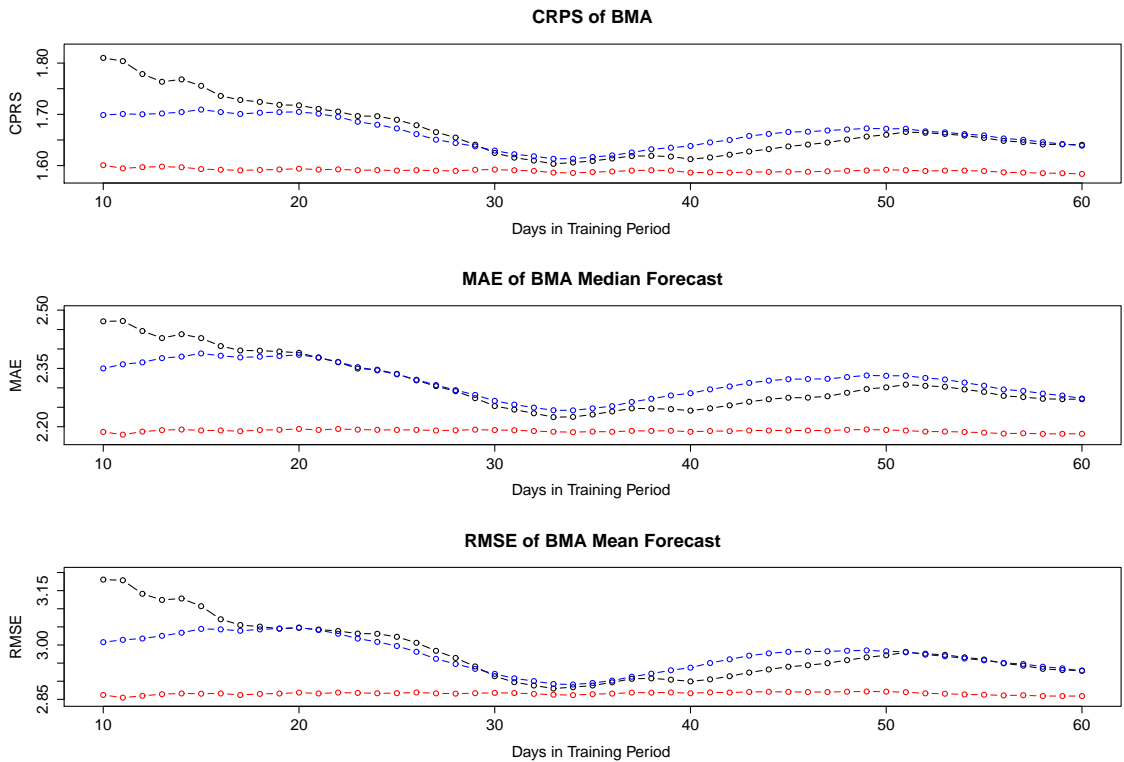,height=15cm, angle=-90}
\caption{Mean CRPS of BMA predictive distributions, MAE of
  BMA median and RMSE of BMA mean forecasts 
  corresponding to two-group model \eqref{eq:eq3.1} for various training period
  lengths and different types of bias correction (black: linear; blue:
  additive; red: no bias correction).}    
\label{fig:fig4}
\end{center}
\end{figure}
Modeling and analysis of temperature data was performed with the help
of the {\tt
  ensembleBMA} package in R \citep{frgs,frgsb}. As a first step the
length of the 
appropriate training period was determined, then the performances
of the BMA post-processed ensemble forecasts corresponding to models
\eqref{eq:eq3.1} and \eqref{eq:eq3.2} were analyzed.  

\subsection{Training period}
   \label{sec:sub4.1}

Similarly to \citet{rgbp} and \citet{bhn} to determine the length
of the training period to be used we compare the MAE values
of BMA median forecasts, the RMSE values of BMA mean forecasts, the
CRPS values of BMA predictive 
distributions and the coverages and average
widths of  $83.33\,\%$  BMA central prediction intervals for
training periods of 
length \ $10,11, \ldots, 60$ \ calendar days. In order to ensure
the comparability of 
the results we consider verification results from 02.12.2010 to
25.03.2011 (114 days; being after the maximum training period length).

\begin{figure}[t!]
\begin{center}
\leavevmode
\epsfig{file=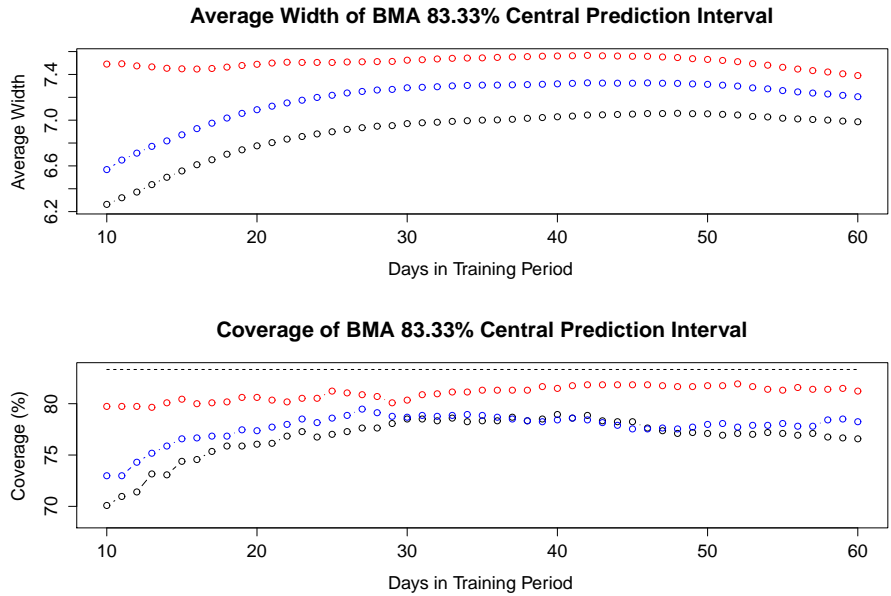,height=12cm, angle=-90}
\caption{Average width and coverage of $83.33\,\%$ 
 BMA central prediction intervals corresponding to two-group model
 \eqref{eq:eq3.1} for various training period 
  lengths and different types of bias correction (black: linear; blue:
  additive; red: no bias correction). Dashed line indicates the
  nominal coverage value.}    
\label{fig:fig5}
\end{center}
\end{figure}
Consider first the two-group model \eqref{eq:eq3.1}. In Figure
\ref{fig:fig4} the CRPS values of BMA predictive distributions, MAE
values of BMA median forecasts and RMSE values of BMA mean forecasts
for different bias correction methods (black: linear; blue:
additive; red: no bias correction) are plotted against the length of
the training period. First of all it is noticeable that the results
are very consistent for each diagnostics, i.e. the curves are similar
for all measures. The best verification scores are obtained without
any bias correction: the CRPS, MAE and RMSE values are rather constant
(on low values) with respect to the length of the training
period, their relative standard deviations are $0.21\,\%$, $0.35\,\%$
and $0.13\,\%$, respectively. In case of linear bias correction the
values of all diagnostics are decreasing with the increase of the
length of the training period reaching their minima around the
32--35 days interval (there is an increase afterwards). Particularly,  
CRPS, MAE and RMSE take their minima at day 33, the corresponding values are
$1.60$, $2.22$ and $2.88$, respectively. For additive bias
correction the patterns of the curves are very similar to that of
the linear case.  The minima of CRPS ($1.61$) and RMSE ($2.89$) are also
reached at 33 days, while MAE takes its minimum of $2.24$
at 34 days (the value at day 33 is very near to this
minimum, in fact that is the second  
smallest one).

Furthermore, Figure
\ref{fig:fig5} shows the average width and coverage of the
$83.33\,\%$ central prediction interval for different bias correction
methods as a
function of the training period length. Concerning the average width the
worst results (largest values) are obtained for 
model \eqref{eq:eq3.1} without bias correction and, similarly to the
previous diagnostics, the curve is rather flat. The linear bias
correction results in the narrowest prediction intervals, the
average width is increasing until around 50 days, then shows
a slight decreasing trend, so up to 48 days, shorter training periods yield
sharper estimates. A similar behaviour can be observed in the case of
additive bias correction (the maximum is reached at around 40 days).

\begin{figure}[t!]
\begin{center}
\leavevmode
\epsfig{file=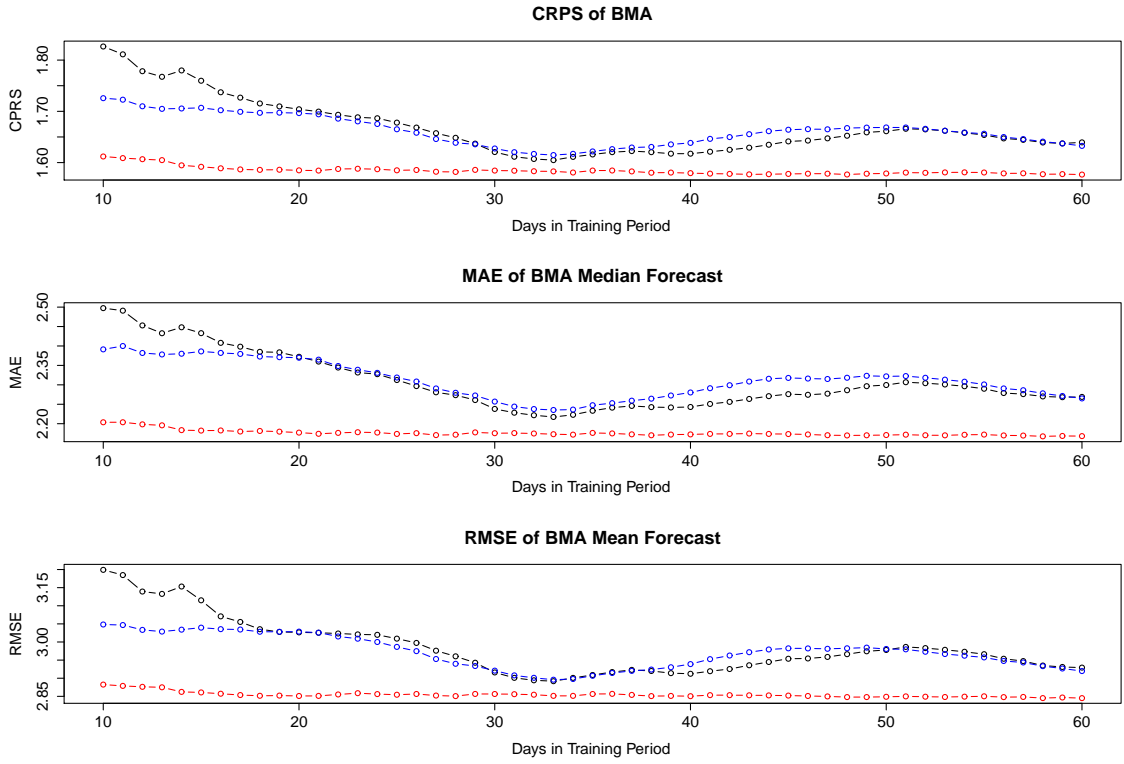,height=15cm, angle=-90}
\caption{Mean CRPS of BMA predictive distributions, MAE of
  BMA median and RMSE of BMA mean forecasts
  corresponding to three-group model \eqref{eq:eq3.2} for various
  training period lengths and different types of bias correction
  (black: linear; blue: additive; red: no bias correction).}   
\label{fig:fig6}
\end{center}
\end{figure}
As far as the coverage values are concerned, unfortunately, none of the
considered models result in coverage reaching the nominal value of 
$83.33\,\%$ (dashed line). However, the case without bias
correction is the best having its maximal value at day 40 with
$81.14\,\%$. Until that day the coverage values are slightly
increasing and then decreasing afterwards (although the curve is rather
flat again). The linear bias correction is starting from the lowest
value and reaching its maxima in the period of 30--40 days (actually,
the maximum is at day 40, but the values at these 10 days are very
similar). The additive bias correction starts higher and reaches its
plateau at around day 30 keeping its rather constant values until the
end of the examined periods. The maximum coverage values of the two
bias corrected cases are slightly below $80\,\%$, the additive one
being better with a small margin. Comparing the average width and
coverage it can be noticed that they have opposite behaviour,
i.e. the average width values favour shorter training periods, while
the coverage figures prefer longer ones. A reasonable compromise
should be found, which is at the range of 20--40 days.  

All in all for all three versions of model \eqref{eq:eq3.1} a
33 days training period seems to be a reasonable choice (particularly
see conclusions based on Figure \ref{fig:fig4}, which are not compromised by
the other two diagnostics at Figure \ref{fig:fig5}).  
\begin{figure}[t!]
\begin{center}
\leavevmode
\epsfig{file=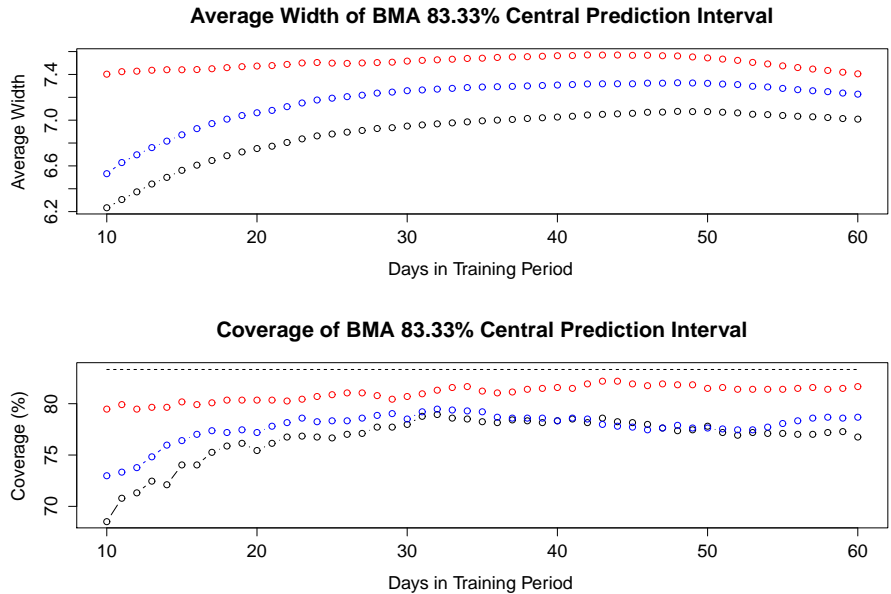,height=12cm, angle=-90}
\caption{Average width and coverage of $83.33\,\%$ 
 BMA central prediction intervals 
  corresponding to three-group model \eqref{eq:eq3.2} for various
  training period lengths and different types of bias correction
  (black: linear; blue: 
  additive; red: no bias correction). Dashed line indicates the
  nominal coverage value.}    
\label{fig:fig7}
\end{center}
\end{figure}

The same conclusion can be drawn from Figures \ref{fig:fig6} and
\ref{fig:fig7} for the variants of the three-group model
\eqref{eq:eq3.2}. The overall behaviour of the different systems for
the various diagnostics is very similar to that for the two-group
model. The no bias correction option provides the best
results all over the time periods in 4 of the 5 diagnostics (the
exception is the average width).  The additive bias correction
slightly outperforms the linear one (except in terms of average
width). In terms of specific values the minima for CRPS, MAE and RMSE,
respectively, are reached again at day 33 
for the linear and additive bias corrections. Regarding the average
width the values are increasing until day 30 and then they are
rather constant until day 50 and decreasing afterwards. For the
coverage the period of highest values are 
between days 30 and 40 (having another maximum of similar value for
the additive bias 
correction at the end of the period). Hence, the training period
proposed for the two-group model can be kept for the three-group
model as well, so we suggest the  
use of a training period of length 33 days for all the investigated
BMA models.

\begin{figure}[t!]
\begin{center}
\leavevmode
\epsfig{file=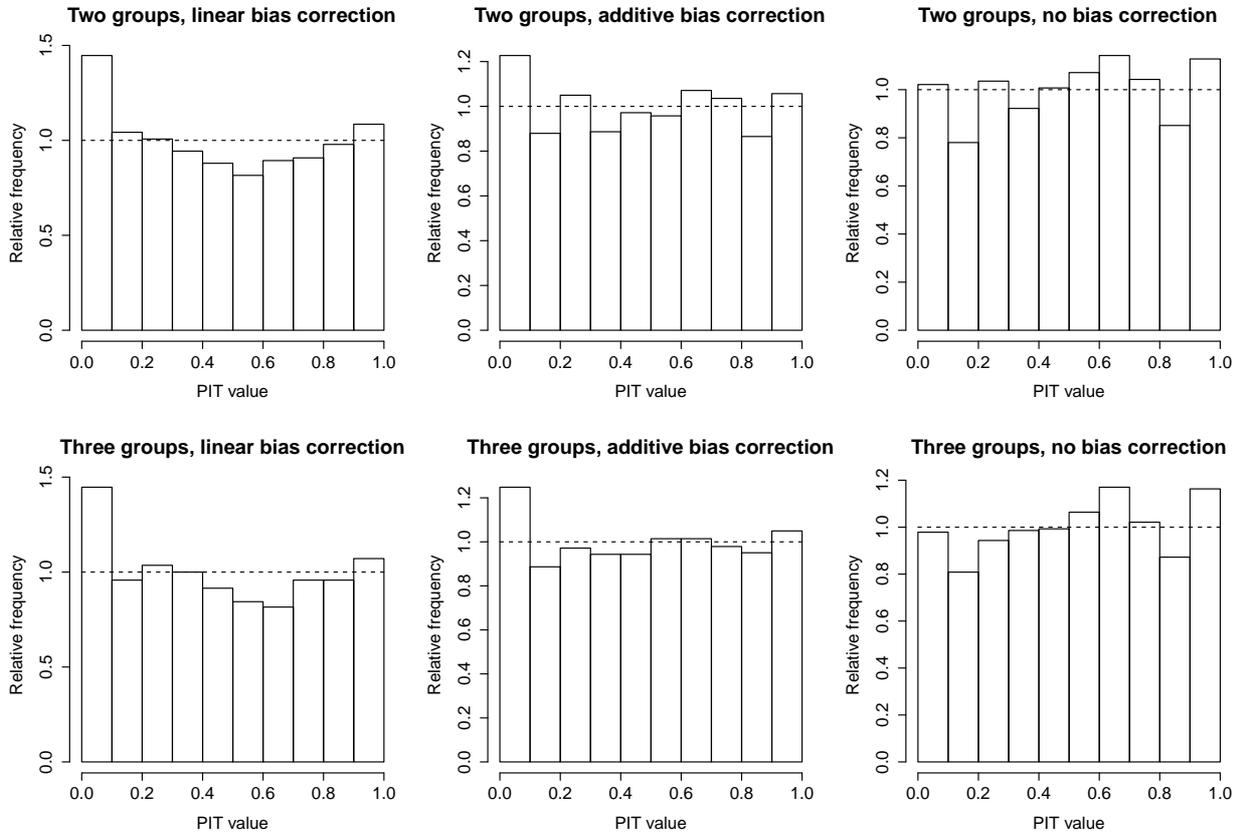,height=16.5cm, angle=-90}
\caption{PIT histograms for BMA post-processed forecasts using two-group
  \eqref{eq:eq3.1} and three-group \eqref{eq:eq3.2} models.}   
\label{fig:fig8}
\end{center}
\end{figure}  
 
\subsection{Ensemble calibration using BMA post-processing}
\label{sec:sub4.2}

According to the results of the previous subsection, to test the performance
of BMA post-processing on the 11 member ALADIN-HUNEPS ensemble we use a
training period of 33 calendar days. In this way ensemble members,
validating observations and BMA models are available for the period
04.11.2010 -- 25.03.2011 (just after the 33 days training period
having 141 calendar days, since on 20.11.2010 all
ensemble members are missing).

To get a first insight about the calibration of BMA post-processed
forecasts we consider probability integral transform (PIT)
histograms. The PIT is 
the value of the predictive cumulative distribution evaluated at
the verifying observations \citep{rgbp}, which is providing a good
measure about the possible improvements of the under-dispersive
character of the raw ensemble.  The closer the histogram is to the
uniform distribution, the better the calibration is. 
In Figure \ref{fig:fig8} the PIT histograms corresponding to all three
versions of  two- and three-group BMA models
\eqref{eq:eq3.1} and \eqref{eq:eq3.2} are displayed. A comparison to
the verification rank histogram of the raw ensemble (see  Figure
\ref{fig:fig1}) shows that post-processing significantly improves the
statistical 
calibration of the forecasts. However, in case of linear bias correction
both PIT histograms are slightly under-dispersive, while for models
with additive bias correction and without bias correction one can
accept uniformity. The latter statement can be also derived from Table
\ref{tab:tab1} where the significance levels of Kolmogorov-Smirnov tests
for uniformity of the PIT values are listed. Based on these tests it
can be concluded that the additive bias correction produces the best
PIT histograms (the no bias correction case is just slightly worse).   

\begin{table}[b!]
\begin{center}
\begin{tabular}{|l|c|c|c|} \hline
\multicolumn{1}{|l|}{}&\multicolumn{3}{|c|}{Bias correction} \\ \cline{2-4}
&linear&additive&none \\ \hline
BMA model \eqref{eq:eq3.1}&$3.7\times 10^{-5}$&$0.25$&$0.20$\\
BMA model \eqref{eq:eq3.2}&$2.9\times 10^{-4}$&$0.22$&$0.06$ \\ \hline 
\end{tabular} 
\caption{Significance levels of Kolmogorov-Smirnov tests for
  uniformity of PIT 
  values corresponding to models \eqref{eq:eq3.1} and
  \eqref{eq:eq3.2}.} \label{tab:tab1}      
\end{center}
\end{table}

\begin{table}[t!]
\begin{center}
\begin{tabular}{|l|c|c|c|c|c|c|c|c|c|c|c|} \hline
\multicolumn{1}{|c|}{}&\multicolumn{1}{|c|}{}&
\multicolumn{1}{|c|}{Mean CRPS}&
\multicolumn{2}{|c|}{MAE}&
\multicolumn{2}{|c|}{RMSE}\\ \cline{4-7}
&Bias corr.&&median&mean&median&mean
\\ \hline 
&linear&$1.54$&$2.15$&$2.15$&$2.77$&$2.77$
\\ 
BMA model
\eqref{eq:eq3.1}&additive&$1.52$&$2.11$&$2.12$&$2.75$&$2.75$
\\ 
&none&$1.52$&$2.10$&$2.11$&$2.75$&$2.75$
\\ \hline
&linear&$1.54$&$2.14$&$2.14$&$2.77$&$2.78$\\
BMA model
\eqref{eq:eq3.2}&additive&$1.52$&$2.10$&$2.11$&$2.75$&$2.75$
\\
&none&$1.52$&$2.09$&$2.10$&$2.74$&$2.74$
\\\hline 
Raw ensemble&&$1.73$&$2.10$&$2.07$&$2.76$&$2.72$\\ \hline
\end{tabular} 
\caption{Mean CRPS of probabilistic, MAE and RMSE of
median/mean forecasts.} \label{tab:tab2}
\end{center}
\end{table}
In Table \ref{tab:tab2} verification measures of probabilistic
and point forecasts calculated using BMA models
\eqref{eq:eq3.1} and \eqref{eq:eq3.2} are compared to the
corresponding scores for the raw ensemble. One can observe that in all cases
there is a significant improvement in the value of the mean CRPS, but
the accuracy of the post-processed point forecasts is equal to or even
slightly worse than the 
accuracy of the corresponding quantities derived from the raw
ensemble. It can be further mentioned that the best scores are
obtained when no bias correction is applied and generally the
three-group model is better than the two-group one. 

\begin{table}[b!]
\begin{center}
\begin{tabular}{|l|ccc|ccc|c|} \hline
\multicolumn{1}{|c|}{}&\multicolumn{3}{|c|}{BMA model \eqref{eq:eq3.1}}&
\multicolumn{3}{c|}{BMA model \eqref{eq:eq3.2}}&
\multicolumn{1}{c|}{Raw ensemble} \\ \hline
Bias correction&linear&additive&none&linear&additive&none& \\ \hline
Average Width
&$6.63$&$6.92$&$7.22$&$6.62$&$6.91$&$7.21$&$2.38$\\ 
Coverage ($\%$)&$77.7$&$79.9$&$81.6$&$77.9$&$80.1$&$
81.8$&$37.4$ \\ \hline  
\end{tabular} 
\caption{Average width and coverage of $83.33\,\%$ central
prediction intervals.} \label{tab:tab3}    
\end{center}
\end{table}

Table \ref{tab:tab3} gives the coverage and average width of the
$83.33\,\%$  central prediction interval calculated using models
\eqref{eq:eq3.1} and \eqref{eq:eq3.2}, and the corresponding measures
calculated 
from the raw ensemble. As before, in the latter case the ensemble of 
forecasts corresponding to a given location and time is considered as a
statistical sample. The BMA prediction intervals calculated from
all models are nearly three times as wide as the corresponding
intervals of the raw ensemble. This comes from the small dispersion of
the raw ensemble, see the verification rank histogram of Figure
\ref{fig:fig1}. It means that although the raw ensemble looks rather
sharp in terms of average width, it cannot be correct, since
the ensemble does not provide sufficient spread in describing the true
variability of the possible atmospheric states. Concerning calibration
one can observe that the coverage of all 
BMA central prediction intervals is much closer to the nominal coverage
than the coverage of the central prediction interval calculated from the
raw ensemble, which is rather poor. 
The diagnostics based on Table \ref{tab:tab3} also confirm that model
\eqref{eq:eq3.2} distinguishing 
three exchangeable groups of ensemble forecasts slightly outperforms
model \eqref{eq:eq3.1} and the best results are obtained without bias
correction. The additive bias correction is slightly better than the
linear one. All these results are in good agreement with the ones
obtained during the determination of the training period length.  

\begin{figure}[t!]
\begin{center}
\leavevmode
\epsfig{file=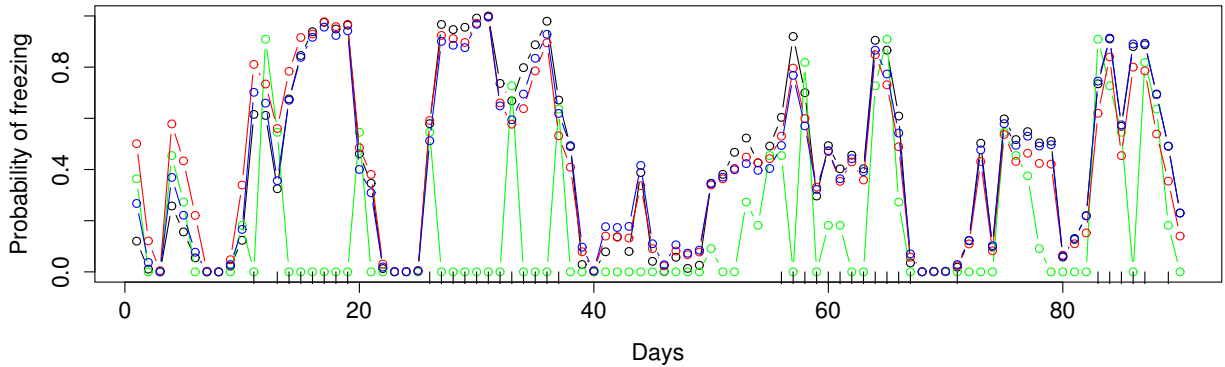,height=6.2cm}
\caption{Probabilities of freezing in Debrecen in the winter period
  01.12.2010 -- 28.02.2011 calculated using the three-group model
  \eqref{eq:eq3.2} with different bias correction methods (black: linear; blue:
  additive; red: no bias correction) and the raw ensemble
  (green). Ticks indicate days with observed temperature below 273 K.}   
\label{fig:fig9}
\end{center}
\end{figure}

\begin{figure}[b!]
\begin{center}
\leavevmode
\epsfig{file=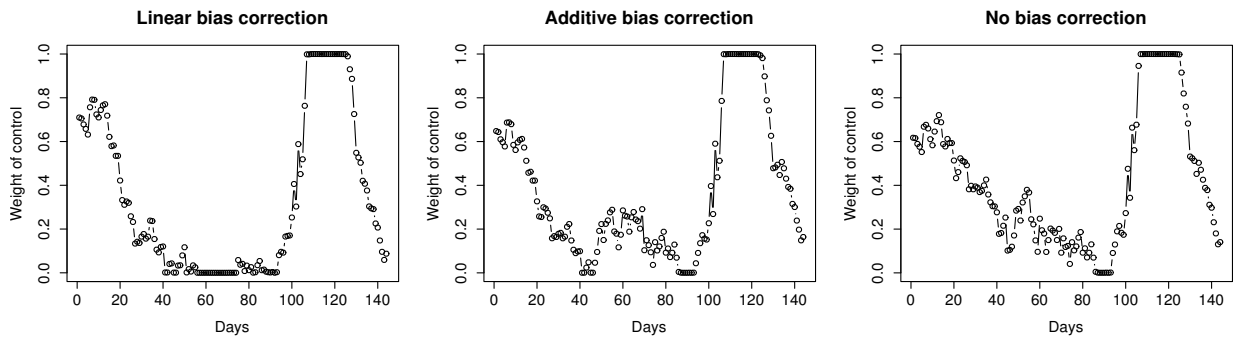,height=4.8cm}
\caption{BMA weights of the control member of the two-group model
  \eqref{eq:eq3.1}.}   
\label{fig:fig10}
\end{center}
\end{figure}

We have decided to consider also other aspects of the probability
density function of the raw and calibrated ensembles. On Figure
\ref{fig:fig9} probabilities of freezing (i.e. temperature forecast
below zero) in the city of Debrecen in the
winter period 01.12.2010 -- 28.02.2011 calculated using the three group model
\eqref{eq:eq3.2} with different bias correction methods and the raw
ensemble  are
 plotted, while ticks indicate the days with observed temperature
 below zero in Celsius (below 273 K). 
This figure immediately reveals that contrary to the observations the
raw ensemble mostly does not indicate any freezing probability (no
members have below-zero temperature forecast). This weakness is
dramatically improved at the calibrated ensembles, where probabilities
have been significantly increased, consequently the calibrated
ensembles have an enhanced prognostic skill than it is the case for
the raw ensemble. This improvement is indeed vital for the operational
application of the ensemble forecasts. 
\begin{table}[t!]
\begin{center}
\begin{tabular}{|l|c|c|c|c|c|c|c|c|c|c|c|} \hline
\multicolumn{1}{|c|}{}&
\multicolumn{1}{|c|}{Control}&
\multicolumn{10}{|c|}{Exchangeable members}
\\ \cline{2-12}
&$f_c$&$f_{\ell,1}$&$f_{\ell,2}$&$f_{\ell,3}$&$f_{\ell,4}$&$f_{\ell,5}$&$f_{\ell,6}$&$
f_{\ell,7}$&$f_{\ell,8}$&$f_{\ell,9}$&$f_{\ell,10}$ \\ \hline
MAE&$1.80$&$1.98$&$1.97$&$2.09$&$2.02$&$2.08$&$1.95$&$2.04$&$
1.85$&$1.84$&$1.96$\\
RMSE&$2.28$&$2.50$&$2.47$&$2.56$&$2.49$&$2.59$&$2.39$&$2.56$&$
2.33$&$2.40$&$2.43$\\ \hline
\end{tabular} 
\caption{MAE and RMSE of the control and exchangeable ensemble
  forecasts for the period  18.02.2011 -- 08.03.2011.} \label{tab:tab4}
\end{center}
\end{table}

\begin{table}[b!]
\begin{center}
\begin{tabular}{|l|c|c|c|c|c|c|c|c|c|c|c|} \hline
\multicolumn{1}{|c|}{}&
\multicolumn{1}{|c|}{Control}&
\multicolumn{10}{|c|}{Exchangeable members}
\\ \cline{2-12}
&$f_c$&$f_{\ell,1}$&$f_{\ell,2}$&$f_{\ell,3}$&$f_{\ell,4}$&$f_{\ell,5}$&$f_{\ell,6}$&$
f_{\ell,7}$&$f_{\ell,8}$&$f_{\ell,9}$&$f_{\ell,10}$ \\ \hline
MAE&$2.62$&$2.44$&$2.75$&$2.55$&$2.75$&$2.52$&$2.55$&$2.63$&$
2.67$&$2.81$&$2.67$\\
RMSE&$3.17$&$3.00$&$3.31$&$3.07$&$3.29$&$3.02$&$3.09$&$3.16$&$
3.28$&$3.37$&$3.29$\\ \hline
\end{tabular} 
\caption{MAE and RMSE of the control and exchangeable ensemble
  forecasts for the period  29.12.2010 -- 16.01.2011.} \label{tab:tab5}
\end{center}
\end{table}
Figure \ref{fig:fig10} shows the BMA weights of the control member \
$\omega$ \ in  three variants of the two-group model
\eqref{eq:eq3.1}. In case of linear bias correction we have a real
mixture in $63.19\,\%$ of the forecasted days (none of the groups has
a weight which is almost 1, e.g. above $0.99$), while for additive
bias correction and for model \eqref{eq:eq3.1} without bias correction
this rate equals  $79.17\,\%$  and $81.25\,\%$, respectively. The
values of \ $\omega$ \ that are close to 1 in all three cases
correspond to a time interval 18.02.2011 -- 08.03.2011,  when the control
member of the ensemble gives much
better forecasts than the ten exchangeable ensemble members (this can
be clearly seen from Table \ref{tab:tab4} where the MAE and RMSE
values of the particular ensemble members are given for the above 
mentioned period). This special behaviour is therefore correctly
diagnosed by the calibration with the provision of large weights for
the control member.  Furthermore, for model \eqref{eq:eq3.1} with linear bias
correction there is a long period between 29.12.2010 and 16.01.2011
where \ $\omega <0.01$ (which means that the control is practically
not used for the calibration), but in the other two cases this phenomenon
practically disappears.  The average MAE and RMSE values for that
period are also displayed for each ensemble member (see Table
\ref{tab:tab5}). It can be seen that the control member is not really
worse than the other ones (it outperforms around half of the 
members), which is correctly ``recognized'' by the additive and no bias
correction techniques. 

\begin{figure}[t!]
\begin{center}
\leavevmode
\epsfig{file=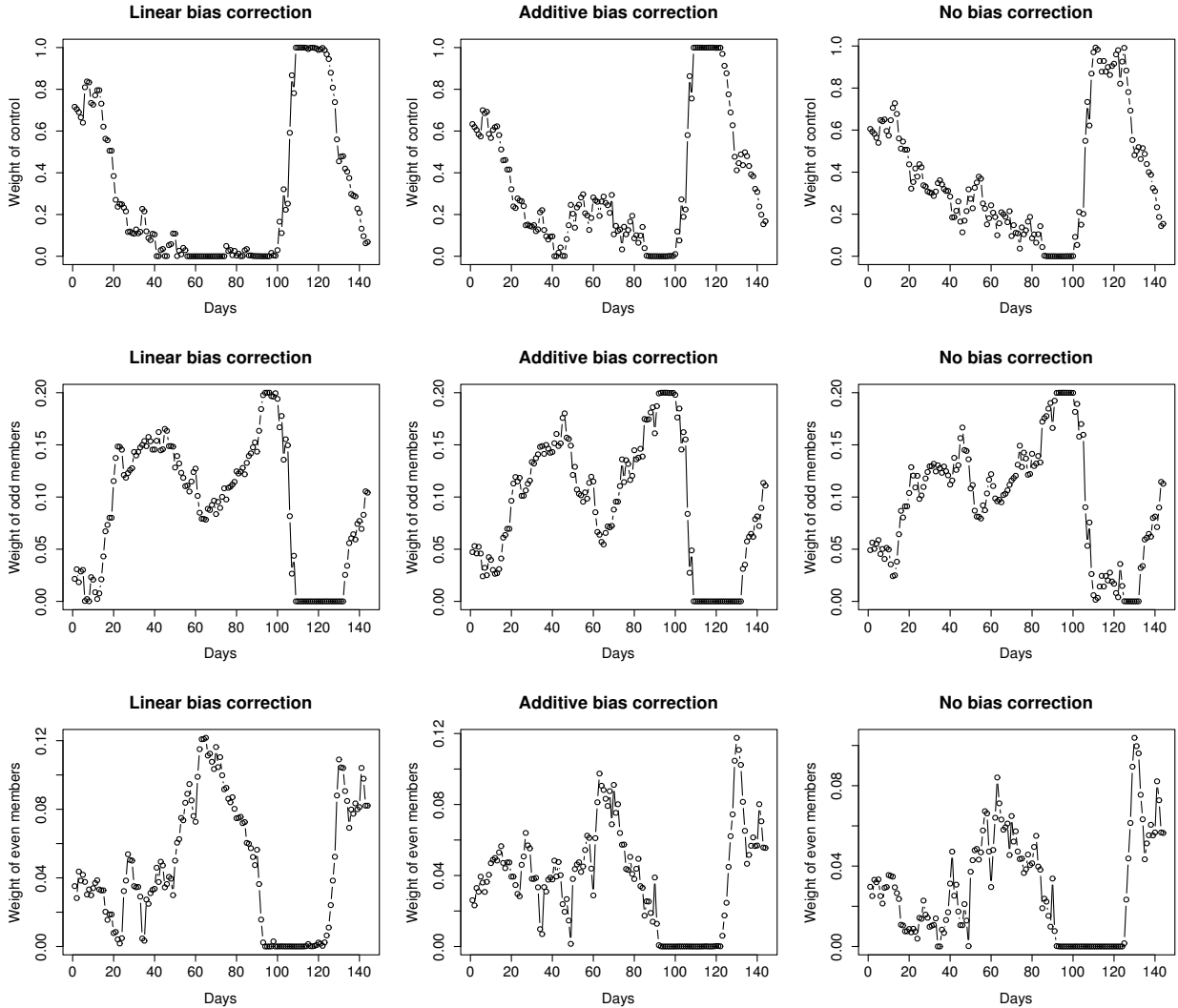,height=14.4cm}
\caption{BMA weights of the control (upper panels) and of the odd
  (middle panels) and even (lower panels) ensemble
  members of the three-group model \eqref{eq:eq3.2}.}   
\label{fig:fig11}
\end{center}
\end{figure}
Finally, Figure \ref{fig:fig11} shows the BMA weights of the control
and of the odd and even ensemble members corresponding to the three
variants of the three-group model \eqref{eq:eq3.2}. In case of linear
and additive bias correction on a part of the problematic period 18.02.2011 --
08.03.2011 (between 20.02.2011 and 05.03.2011) the weight \ $\omega_c$
\ of the control is still close to $1$ (greater than $0.98$), while
for the model without bias correction this happens only on four days. In
the remaining cases ($89.58\,\%$, $90.28\,\%$ and $97.22\,\%$ of the days,
respectively) we have real mixtures of normal distributions (it is
particularly remarkable for the no bias correction). On the other hand
for the period, when the control gets zero weight in the linear bias
correction it is again (correctly) partly disappearing in the additive
and no bias correction cases.

\section{Conclusions}
   \label{sec:sec5}

In the present study the BMA ensemble post-processing method is applied 
to the 11 member ALADIN-HUNEPS ensemble of the HMS to obtain 42 hour
calibrated 
predictions for 2m temperature. Two different BMA models are
investigated, one assumes two groups of exchangeable members (control
and forecasts from perturbed initial conditions), while the other
considers three (control and forecasts from perturbed initial
conditions with positive and negative perturbations). In both cases
three different treatments of bias are considered (linear and additive
bias correction and no bias correction at all) and for all models  
a 33 days training period is suggested. The comparison of the
raw ensemble and of the probabilistic forecasts shows that the mean
CRPS values of BMA post-processed forecasts are considerably lower
than the mean 
CRPS of the raw ensemble, while there is no big change in the MAE and
RMSE values of BMA point 
forecasts (median and mean) compared to the MAE and RMSE of the
ensemble median and of the ensemble mean. This latter fact might mean
that although the spread of the raw ensemble was too small, the median
and mean of the point forecasts were sufficiently correct not having
too much space for further improvement. It is remarkable to notice
that the real probabilistic forecasts as demonstrated by the
probability of freezing for Debrecen have been significantly improved.  
For models with additive
bias correction and without bias correction the coverage of the
$83.33\,\%$ central prediction interval is rather close to the
nominal value, while their PIT values fit the uniform
distribution. From the six competing post-processing methods the
overall performance of the three-group model without bias
correction seems to be the best, while models with linear bias correction
give the worst results.

Finally, we conclude that BMA post-processing of the ALADIN-HUNEPS
temperature ensemble forecasts significantly improves the calibration
and the probabilistic forecasts, however 
no significant changes are found in accuracy of point forecasts. The first
two aspects encourage the operational application of the method, while
the third one calls for some further investigations to see whether
improvements can be made for the point forecasts as well.  

\bigskip
\noindent
{\bf Acknowledgments.} \  \ Research was supported by 
the Hungarian  Scientific Research Fund under Grant No. OTKA NK101680
and by the T\'AMOP-4.2.2.C-11/1/KONV-2012-0001 
project. The project has been supported by the European Union,
co-financed by the European Social Fund.  The authors are indebted to Tilmann
Gneiting for his useful suggestions and remarks and to M\'at\'e Mile and 
Mih\'aly Sz\H ucs from the HMS for providing the data.

\end{document}